\documentstyle[epsfig]{aipproc2}
\begin{document}    
\renewcommand{\topfraction}{0.95}
\renewcommand{\bottomfraction}{0.95}
\renewcommand{\textfraction}{0.05}
\title{Stability and
Production of Superheavy Nuclei}
 
\author{Peter M\"{o}ller$^{*,{\dagger}}$ and J. Rayford Nix$^{\dagger}$}
\address{$^*$P. Moller Scientific Computing and Graphics, Inc.,\
P.\ O.\ Box 1440, Los Alamos, New Mexico 87544, USA\\
$^{\dagger}$Theoretical Division,
Los Alamos National Laboratory, Los Alamos, New Mexico 87545, USA}
 
\maketitle          
 
\begin{abstract}    
Beyond uranium heavy elements rapidly become
increasingly unstable with respect to spontaneous fission
as the  proton number $Z$ increases,
because             
of the disruptive effect of the long-range Coulomb force. However,
in the region just beyond $Z=100$ magic proton and neutron numbers
and the associated shell structure
enhances nuclear stability sufficiently to allow observation
of additional nuclei. Some thirty years
ago it was speculated that an island of spherical, relatively
stable superheavy nuclei would exist near the next doubly magic
proton-neutron combination
beyond $^{208}$Pb, that is, at proton number $Z=114$ and neutron number
$N=184$.            
Theory and experiment now show that there also exists a rock of stability
in the vicinity of $Z=110$ and $N=162$  between the actinide
region, which previously was the end of the peninsula of known elements,
and the predicted island of spherical superheavy nuclei slightly
southwest of the magic numbers
$Z=114$ and $N=184$. We review here the stability properties of the
heavy region of nuclei.
 
Just as the decay properties of  nuclei in the heavy region
depend strongly on shell structure, this structure also
dramatically affects the fusion entrance channel. The six most recently
discovered new elements were all formed in cold-fusion reactions.
We discuss here the effect of the doubly magic structure of the
target in cold-fusion reactions on the fusion barrier and on
dissipation.        
\end{abstract}      
 
\section*{Introduction}
It was predicted more than 30 years ago  that the next doubly magic
nucleus beyond $^{208}_{\phantom{0}82}$Pb$_{126}$ is
$^{298}$114$_{184}$ 
and that nuclei on an island of
superheavy nuclei in its vicinity   have
half-lives of up to perhaps several billion
years [1--7].       
In Fig.~\ref{fig01moln} we show the calculated enhancement to binding due to microscopic
effects [8]         
for nuclei throughout the periodic system. The two nuclei
$^{266}$Pb and $^{238}$Hf
are so near the neutron drip line that one can anticipate that they
will never be observable. Of the remaining doubly magic regions
only the recently observed region near $^{272}$110 consists of
nuclei deformed in their ground states. The only spherical doubly magic
region that realistically remains to be observed is therefore the  island
of spherical superheavy elements currently predicted to occur near $^{292}$114.
\begin{figure}[t]   
\begin{center}      
\vspace{3.4in}
\caption[fig01mol]{Calculated microscopic enhancement  to binding
for nuclei throughout the periodic system.}
\label{fig01moln}   
\end{center}        
\end{figure}        
Although no nuclei on
this superheavy island have been observed so far,
six new elements with proton numbers $Z=107$--112 have been discovered
between this superheavy island and the previously heaviest-known elements
at the edge of the actinide region. The heaviest of the new
elements are localized in the vicinity of a rock
of {\it deformed\/} shell-stabilized nuclei near proton number $Z=110$
and neutron number $N=162$.
We show in Fig.~\ref{fig02moln}
the location of the currently  heaviest known nucleus $^{277}$112 and
its                 
observed $\alpha$-decay products
superimposed on calculated microscopic corrections for nuclei
throughout the periodic system.
\begin{figure}[b]   
\begin{center}      
\vspace{-0.2in}     
\vspace{3.56in}
\vspace{-0.3in}     
\caption[fig02mol]{The heaviest known element $^{277}$112
and its $\alpha$-decay products as dots superimposed on
calculated  microscopic corrections for nuclei throughout
the periodic \vspace{-0.5in}system.}
\label{fig02moln}   
\end{center}        
\end{figure}        
In the following we discuss what we have learned experimentally and
theoretically about (1) the stability properties of the heaviest nuclei and (2)
the cold-fusion reaction mechanism that  has been so essential
in the discovery of the heaviest elements.
A few aspects of hot fusion will also be reviewed.
 
\section{Models}    
 
In the study of nuclear structure  a substantial number of different models
are used, many of which are discussed at this conference.
The significance of  these models and their relation to each other
is made more clear if we observe that
it is possible to put the nuclear models used into one of the
 groups below:      
\begin{description} 
\item[1.]  Models where the physical quantity of interest is given
    by an expression such as a polynomial or a more
    general algebraic expression. The parameters are usually
    determined by adjustments to experimental data. Normally models of this
    type describe  only a single nuclear property.
No nuclear wave functions are obtained in these models.
\begin{figure}[b]   
\begin{center}      
\vspace{-0.2in}     
\vspace{3.2in}
\caption[fig03mol]{ 
Comparison between energy releases $Q_{\alpha}$ obtained in the FRDM (1992),
ETFSI-1 (1992) model, and FDSM (1992) and recent experimental data
for the heaviest known element. Two different decay chains were observed
experimentally.     
}                   
\label{fig03moln}   
\end{center}        
\end{figure}        
An example of       
a model of this type is the original Bethe-Weizs\"{a}cker macroscopic mass
model. A nice feature of this macroscopic mass model is that it can be
generalized, without any additional parameters, to describe macroscopic
fission barriers. This  generalization property is unusual for
models in this category.

\item[2.]           
Models that are claimed to be based on  microscopic equations
with realistic two-body interactions
but which utilize   
so many approximations
that the end result is again some algebraic expression with
parameters that are adjusted to
experimental data. No microscopic differential equations
are actually solved.
\begin{figure}[t]   
\begin{center}      
\vspace{-0.2in}     
\vspace{3.0in}
\caption[fig04mol]{Nuclear shapes leading from the ground state
to elongated and compact scission shapes. The corresponding potential-energy
surface for $^{264}$Fm is shown in Fig.~\ref{fig05moln}. The quantity
$R_0$ is the radius of the original spherical nucleus.
}                   
\label{fig04moln}   
\end{center}        
\end{figure}        
Examples of models of this type are the Duflo mass model [9] and the
fermion dynamical-symmetry model [10] which is
applied to nuclear mass calculations and some other calculations.
 
\item[3.]  Models that use an effective nuclear
interaction and {\it  actually solve\/} the
resulting  microscopic  quantum-mechanical equations,
for example a Schr\"{o}dinger or a Dirac equation. The solutions provide {\it nuclear
wave functions\/} which allow {a vast number} of nuclear properties to be
modeled within {\it a single\/}  framework.
\begin{figure}[b]   
\begin{center}      
\vspace{-0.2in}     
\vspace{3.0in}
\caption[fig05mol]{Potential-energy surface leading from the ground state
to elongated and compact scission shapes. The corresponding shapes are
shown in Fig.~\ref{fig04moln}.
}                   
\label{fig05moln}   
\end{center}        
\end{figure}        
Most models of this type that are currently used fall into two subgroups
depending on the type of wave function
and the type of microscopic interaction  used:
\begin{description} 
\item[3a] Single-particle models that use a simple central potential
with additional residual interactions such as the two-body pairing interaction.
\begin{figure}[b]   
\begin{center}      
\vspace{-0.2in}     
\vspace{3.5in}
\caption[fig06mol]{ 
Proton single-particle level diagram for merging nuclei in an
asymmetric heavy-ion collision leading to the heaviest known nucleus.
The intersecting-sphere parameterization is used for the entire path
from touching spheres to a single sphere. Thus, the level-diagram path
does not pass exactly through the calculated ground-state shape.
The asymmetric configurations in the entrance channel lead to a mixing
of states with odd and even parity. The magic-fragment gaps associated
with the initial entrance-channel configuration remain far inside
the touching point, to about $r/R_0=1.15$, somewhat outside the maximum
in the fusion barrier.
}                   
\label{fig06moln}   
\end{center}        
\end{figure}        
The  Schr\"{o}dinger equation is solved in a single-particle approximation
and additional two-body interactions are treated in the BCS, Lipkin-Nogami,
or RPA approximations, for example. Wave functions are obtained, which
allows a large number of nuclear-structure features to be predicted,
such as transition rates within or
between rotational bands or beta-decay transition rates.
 
For the calculation of the total nuclear energy it is not possible in
the single-particle model
to obtain the nuclear ground-state energy
as $E=\; <\!\Psi_0|H|\Psi_0\!>$,  where $\Psi_0$ is
the ground-state nuclear wave function. To obtain the
nuclear potential energy as a function of shape one combines
the single-particle model
\begin{figure}[b]   
\begin{center}      
\vspace{-0.2in}     
\vspace{4.1in}
\caption[fig07mol]{ 
Total, adiabatic, and macroscopic fusion barriers for the cold-fusion reaction
$^{70}$Zn~+~$^{209}$Bi~$\rightarrow$~$^{279}$113
and fission barrier corresponding to spontaneous fission  from
the ground state.   
}                   
\label{fig07moln}   
\end{center}        
\end{figure}        
with a macroscopic model, which leads to the macroscopic-microscopic model in which the
energy is calculated as a sum of a microscopic correction obtained
from calculated single-particle levels by use  of the Strutinsky method
and a macroscopic energy.
\item[3b] Hartree-Fock-type models in which the postulated effective
interaction is of two-body type and the wave function is
an antisymmetrized Slater determinant. In other respects, these models
have many similarities to those in {\bf 3a}, with the exception that it is possible to
obtain the nuclear ground-state energy as \mbox{$E=\;<\!\Psi_0|H|\Psi_0\!>$}.
\end{description}   
\end{description}   
We sometimes hear proponents of models of type {\bf 3b} refer to models of type
{\bf 3a} as ``macroscopic.'' This is clearly inaccurate since
 
\begin{description} 
 
\item[a]both models 
solve microscopic wave equations. The difference between the two
approaches lies in the type of interaction and type of wave function used.
 
\item[b] to date most if not all important new insight into {\it microscopic}
nuclear structure has been provided by models in category {\bf 3a}.
\begin{figure}[b]   
\begin{center}      
\vspace{-0.2in}     
\vspace{3.5in}
\caption[fig08mol]{ 
Seven touching configurations for heavy-ion collisions with spherical
projectiles and general targets.
}                   
\label{fig08moln}   
\end{center}        
\end{figure}        
For example, reasonably accurate ground-state deformations and masses,
nuclear level structure, including spins and parities, and the mass asymmetry
of the fission saddle point were first obtained in this approach.
\begin{figure}[b]   
\begin{center}      
\vspace{-0.2in}     
\vspace{3.34in}
\caption[fig09mol]{ 
Five touching configurations for heavy-ion collisions with prolate,
negative-hexadecapole projectiles and targets.
}                   
\label{fig09moln}   
\end{center}        
\end{figure}        
The first two model categories are only able to parameterize or polynomialize
data by use of expressions that normally contain a vast number of parameters
and consequently reproduce experimental data used in the adjustment well, but fail
catastrophically for data not used in the adjustments or for new data.
\begin{figure}[b]   
\begin{center}      
\vspace{-0.2in}     
\vspace{4.2in}
\caption[fig10mol]{ 
Calculated potential-energy surface in units of MeV for the reaction
$^{48}$Ca~+~$^{244}$Pu. The energy in the medium-gray area outside the
target nucleus in the center was not calculated, because the points
in this region correspond to points inside the touching configuration.
Note the ridge with saddle points and peaks around the
target.             
}                   
\label{fig10moln}   
\end{center}        
\end{figure}        
Models in category {\bf 3b} are expected to {\it in principle} be
more accurate than models in   category {\bf 3a}, because
 the wave function is more realistic and  more realistic
effective-interactions
 can be used. However, two problems
remain today: what effective two-body interaction is {\it realistic},
in the sense that it
will yield more accurate results than the well-studied and
well-optimized single-particle effective interactions, and what are the optimized
parameter values of such a realistic two-body interaction?
\end{description}   
 
Let us further emphasize that there is no ``correct'' model in nuclear physics.
Modeling of nuclear physics involves simplifying the true forces and equations
with the goal to obtain a formulation
that can be solved in practice, but that ``retains the essential features''
of the true system under study. How to obtain an effective force from
the true force is not well-defined. What we mean by ``retains the essential
features'' depends on the circumstances. Simply speaking, it means that
it retains sufficiently much of the true system that we can learn something
from the simplified model. Bearing this in mind, it is clear that the microscopic
single-particle models have been enormously successful over the
years, since we have learned so much from them.
 
In our brief overview here we will discuss
what has been learned about the
stability of the heaviest elements
from calculations based on a macroscopic-microscopic model
with a realistic diffuse-surface
folded-Yukawa single-particle potential as a starting
point for the microscopic term and a generalized droplet model
with a Yukawa-plus-exponential potential for the nuclear energy in the
macroscopic term. The model has been described in sufficient detail
elsewhere [8,11--14].
For a review of other results  we refer to the many other interesting
talks at this conference and to a recent review [15].

\section{Stability at the end of the periodic system}
 
Whereas predictions of an island of superheavy elements were made
already 30 years ago, the existence of a ``rock'' of relatively long-lived
{\it deformed\/} neutron-deficient shell-stabilized
superheavy nuclei in the vicinity of $Z=110$ and
$N=162$ has been a subject of study only for about 15 years or so.
In our first global nuclear mass calculation in 1981
[13,14],            
which was limited to 4023 nuclei, and which
did not reach the neutron and proton drip lines, part of this rock is
nevertheless clearly visible in the tabulated microscopic corrections.
In Fig.~\ref{fig01moln} we see that the island of superheavy elements
predicted in the mid 1960s is not isolated from the relatively stable
actinide region. Instead, a stabilizing peninsula extends from
the spherical superheavy region towards the actinide region. On this
peninsula two stabilizing ridges corresponding to $N=152$ and $N=162$
are clearly visible.
 
For additional clarity we have included in
Figs.~\ref{fig01moln} and \ref{fig02moln} only even-even nuclei
so that odd-even staggering is removed. The ridge at $N=152$ has long
been connected to the unusually long spontaneous-fission half-lives of
$^{250}$Cf, $^{252}$Fm, and
$^{254}$No [16--18].
 
The peninsula structure extending from the spherical superheavy
island is somewhat similar
to the smaller peninsula extending to the southwest from
the doubly magic $^{208}$Pb. On this peninsula the
most prominent  shell-stabilized ridges occur at $N=102$
and $N=108$, but are less developed than the ridges in the heavy
region.

In Fig.~\ref{fig02moln} we have superimposed the location of the most
recently discovered heavy element $^{277}$112 and
its $\alpha$-decay daughters [19]
on calculated microscopic corrections obtained in the FRDM (1992) [8].
The $\alpha$-decay $Q$-values are plotted in  Fig.~\ref{fig03moln} and
compared to three theoretical
calculations [8,10,20,21].
Two experimental decay chains have been observed. These decay chains
provide for the first time the half-life of a nucleus, $^{271}$Hs,
at the center of the predicted
rock of stability. The measured half-life of about 10~s is in excellent
agreement with the half-life of 630~s predicted by the FRDM (1992)
[20].               
The  agreement between these experimental observations and theoretical
predictions confirms the predictive powers of current nuclear-structure
models              
and represents a triumph for nuclear physics.

\section{The cold-fusion entrance channel}

The six heaviest-known elements were
all produced in cold-fusion reactions with  doubly magic
$^{208}_{\phantom{0}82}$Pb$_{126}$ or near-doubly magic
$^{209}_{\phantom{0}83}$Bi$_{126}$ targets.
The cold-fusion reaction has long been thought to enhance
heavy-element evaporation-residue cross sections primarily because
it leads to compound nuclei of low excitation energy,
which enhances de-excitation by neutron emission relative
to fission. Higher excitation energies would lead to higher
fission probabilities. However, the evaporation-residue cross
section is the product of the
cross section for compound-nucleus formation
and the probability for de-excitation by
neutron emission.   
One may therefore ask if  cold fusion {\it also\/}
enhances the cross section for  compound-nucleus formation.
Because of the low excitation energies in the entrance
channel, the large negative shell correction
associated with target nuclei near the  doubly magic $^{208}$Pb
should be almost fully manifested at touching and slightly inside
touching.           
 
Nuclei near $^{258}$Fm have
already provided important insight into  fragment
shell effects in symmetric fission and
fusion configurations [15,22,23].
At $^{258}$Fm fission becomes symmetric with a very narrow
mass distribution, the kinetic energy of the fragments is
about 35 MeV higher than in the asymmetric fission of $^{256}$Fm,
and the spontaneous-fission half-life is 0.38 ms for $^{258}$Fm compared to
2.86 h for $^{256}$Fm. These features are  well understood
in terms of the macroscopic-microscopic model. Shell effects
associated with division into fragments near $^{132}$Sn
lower the fusion valley at touching by  about 20 MeV
in the most favorable case relative to that in  macroscopic
model. This fragment shell effect remains important
far inside the touching point and results in fission into
the fusion valley with very compact {\it cold\/}
fragments for several fissioning nuclei in the vicinity of $^{258}$Fm.
The maximum effect of fragment shells in symmetric fission occurs in
the fission of the hypothetical nucleus $^{264}$Fm.
In Fig.~\ref{fig04moln} we show a set of shapes leading from
the nuclear ground state towards  elongated scission configurations
in the upper right part of the figure and to a compact scission configuration
in the lower right part of the figure.
The corresponding potential-energy surface calculated in the FRDM (1992)
is shown in Fig.~\ref{fig05moln}. The deep valley to the lower right,
corresponding to cold fission into spherical fragments
or to cold fusion of two $^{132}$Sn nuclei,
is very prominent   
and extends to the inner fission saddle near the ground state.
The deformation coordinates $r$ and $\sigma$ correspond to
the distance between the centers of mass of the two parts of the system
and to the  sum of the root-mean-square extensions along the symmetry axis
of the mass of each half of the system about its center of mass, respectively.

We now show that shell effects
are also very important in the fusion entrance channel in
cold-fusion heavy-ion reactions, which usually involve {\it asymmetric\/}
projectile-target combinations.
In Fig.~\ref{fig06moln}
we show calculated  proton  single-particle levels
for merging, intersecting spheres in terms of the $r$
shape coordinate for the reaction
$^{70}$Zn~+~$^{208}$Pb~$\rightarrow$~$^{278}$112.
This represents the reaction employed to reach the heaviest nucleus
known thus far.     
We note that        
 the magic-fragment gap
combination $28+82=110$ remains far inside the touching point, up
to about $r/R_0=1.15$. The quantity $R_0$ is the radius of the spherical
compound system. Because of the stability of the fragment gaps
during the merging of the two nuclei, excitation should be minimal
until late in the fusion process, which should favor
evaporation-residue formation.
These results are in excellent agreement with the results of
calculations related to the symmetric fission of nuclei near $^{258}$Fm into
symmetric spherical fragments near $^{132}$Sn.
 
To quantitatively study the effect of the persistent
magic-fragment gaps on the fusion barrier as the heavy ions
merge, we have calculated the fusion barrier for intersecting spheres
for the proposed reaction\\
\mbox{ } \hfill $^{70}$Zn~+~$^{209}$Bi~$\rightarrow$~$^{279}$113 \hfill \mbox{ } \\
which is shown in Fig.~\ref{fig07moln}. Just inside
the peak in the fusion barrier at about $r/R_0=1.0$ we have
switched from the intersecting-sphere parameterization
to Nilsson's perturbed-spheroid $\epsilon$ parameterization so that
we accurately obtain the energy of the ground state.
Such a switch is not carried out in the calculation of the level diagram, so
the level-diagram path does not pass exactly through the calculated ground-state
shape.              
The calculated ground-state shape is indicated in Fig.~\ref{fig07moln}.
 We also            
show the touching configuration and one intersecting-sphere
configuration at $r/R_0=1.0$, near  the maximum in the fusion
barrier. The dotted line shows the calculated fission barrier,
for which  we considered  $\epsilon_2$, $\epsilon_4$, and
$\epsilon_6$ shape distortions. The effect of mass asymmetry
on the fission barrier
is expected to be small.
The fusion barrier in the macroscopic FRDM without any shell effects
is given by the short-dashed line.
The touching configuration is indicated by a thin
vertical long-dashed line. The thicker long-dashed line is the calculated
adiabatic barrier without any specialization energy. Despite
the fairly high spin 9/2 of the $^{209}$Bi ground state the
calculated specialization energy is quite low. The adiabatic curve
is  shown only from touching to about $r/R_0=1.0$. However, the specialization
energy is quite low also between $r/R_0=1.0$ and the ground-state shape.
For the cold-fusion reactions that have resulted in the formation of
the elements $Z=107$--112, our calculations show that the incident energy
resulting in  maximum 1n cross section corresponds to an energy just a few MeV
above the calculated total fusion barrier.
 
\section{Superheavy elements by hot-fusion reactions}
 
When hot-fusion reactions are used to produce heavy elements, these
reactions usually involve a deformed actinide target and a spherical,
light projectile. The combination of a spherical
projectile and deformed target is illustrated for various
representative target deformations in Fig.~\ref{fig08moln}.
For such configurations the fusion
barrier is not one-dimensional, but two-dimensional [24,25].
More generally, both the projectile and targets may be deformed. We illustrate
the case of prolate, negative-hexadecapole targets and projectiles
in Fig.~\ref{fig09moln}.
In this case the full characterization of  the
potential between the deformed projectile and
the  deformed target leads to a four-dimensional potential-energy surface.
However, major features of this multi-dimensional potential-energy
surface can be obtained from a calculation of the barrier for the five
limiting configurations shown in Fig.~\ref{fig09moln}.
 
As limiting orientations we consider only situations where
the projectile center is on the $x$, $y$, or $z$ axis of the target and
orientations of the projectile where the projectile symmetry axis is either
parallel to or perpendicular to the target symmetry axis.
Since we restrict ourselves to axial symmetry, configurations with
the projectile center  located on the $x$ or $y$ axis are identical.
If the projectile is located in the equatorial region of
the target it can be oriented in three major orientations,
and if it is located in the polar
region it can be oriented in two major orientations. Thus, for a particular
projectile-target deformation combination there are five
possible limiting configurations. Because compact touching configurations
are thought to favor compound-nucleus formation, a particularly favorable
configuration could be the equatorial-cross configuration, which is the
configuration       
shown to the right in the bottom row of Fig.~\ref{fig09moln}. For
the prolate, negative-hexadecapole targets and projectiles shown in this
figure we call this configuration the ``hugging'' configuration.
Another close-approach configuration would involve an oblate target
and an oblate projectile in a polar-parallel configuration.
The fusion-barrier configurations in deformed heavy-ion collisions are
discussed in greater detail in Ref.~[25].

To reach element $Z=114$ the reaction
$^{48}$Ca~+~$^{244}$Pu~$\rightarrow$~$^{292}$114 is being considered.
In Fig.~\ref{fig10moln} we show our calculated two-dimensional fusion
potential-energy surface for this reaction. The target is centered at
the origin. The fusion potential is shown for locations ($\rho,z)$ of
the center of the projectile, where $\rho$ is the distance from the symmetry
axis $z$. Since the projectile center is some distance away from the
target surface when the target and projectile surfaces touch, the energy
function that we use is not defined inside the gray area
for the separated projectile-target configuration.
It is immediately clear from the figure that there is
a substantial difference between the polar barrier height of 193.3 MeV
and the equatorial barrier height of 208.0 MeV\@.

We have earlier argued that compact hugging configurations are
desirable for evaporation-residue formation [25].
That is, targets and
projectiles should have positive
values of $\epsilon_4$ (negative hexadecapole moments) so that  waistlines
develop.            
This permits a close approach in the equatorial-cross orientation.
Suitable targets would then be nuclei in the rare-earth region
starting at approximately $Z=68$. Corresponding projectiles would
then be in the range $Z=42$--50.
One may, for example, consider a reaction with a spherical projectile:
$^{124}$Sn~+~$^{176}$Yb~$\rightarrow$~$^{300}$120,
for which the equatorial barrier is 378.2 MeV\@.
This means that only two neutrons would be emitted for collisions
at the barrier. An example of a true hugging configuration with
negative-hexadecapole target and projectile shapes is
$^{114}$Cd~+~$^{180}$Hf~$\rightarrow$~$^{294}$120.
Here, the {\it hugging} barrier is 381.8 MeV and
the number of evaporated neutrons is four. The distance between
mass centers for this configuration at touching is 11.08 fm.
Reactions with oblate projectiles exhibit interesting features.
An example is       
$^{116}$Cd~+~$^{180}$Hf~$\rightarrow$~$^{296}$120,
for which the polar-parallel  barrier is
365.6 MeV\@, corresponding to a 2n evaporation process for
collisions at the barrier. The touching distance is
11.91 fm for this orientation. For the equatorial-transverse
orientation the touching distance is only 10.49 fm,
but the barrier is 389.3 MeV for this orientation,
corresponding to the emission of five neutrons.
 
The cold-fusion process between spherical projectiles
and  targets has led to the discovery of six new elements.
Some features of the cold-fusion process are well understood today
but other important aspects remain to be explored. In particular,
we need to understand how  the evaporation-residue cross section
depends on the projectile and target species and on the reaction energy.
The exploration of reactions between spherical or deformed projectiles and deformed targets
leading to heavy elements also presents a  fascinating challenge for
the future. Here we need to understand also the influence of deformation and
relative orientation of the target and projectile
on the evaporation-residue cross section. Also,
in this case one can find fairly cold reactions leading to the evaporation of relatively
few neutrons. In the future, some aspects of these reactions will perhaps
be explored in a radioactive-ion-beam facility with polarized targets.

\newpage            
\begin{center}                                                                  
{\bf References}                                                                
\end{center}                                                                    
\newcounter{bona}                                                               
\begin{list}%
{\arabic{bona})}{\usecounter{bona}                                              
\setlength{\leftmargin}{0.5in}                                                  
\setlength{\rightmargin}{0.0in}                                                 
\setlength{\labelwidth}{0.3in}                                                  
\setlength{\labelsep}{0.15in}                                                   
}                                                                               
\item                                                                           
H.\ W.\ Meldner, unpublished (1965).                                            
                                                                                
\item                                                                           
H.\ W.\ Meldner, Proc.\ Int.\ Symp.\ on why and how to investigate nuclides far 
  off the stability line, Lysekil, 1966, Ark.\ Fysik {\bf 36} (1967) 593.       
                                                                                
\item                                                                           
W.\ D.\ Myers and W.\ J.\ Swiatecki, Ark.\ Fys.\ {\bf 36} (1967) 343.           
                                                                                
\item                                                                           
S.\ G.\ Nilsson, J.\ R.\ Nix, A.\ Sobiczewski, Z.\ {Szyma\'{n}ski}, S.\ Wycech, 
  C.\ Gustafson, and P.\ {M\"{o}ller}, Nucl.\ Phys.\ {\bf A115} (1968) 545.     
                                                                                
\item                                                                           
S.\ G.\ Nilsson, C.\ F.\ Tsang, A.\ Sobiczewski, Z.\ {Szyma\'{n}ski}, S.\       
  Wycech, C.\ Gustafson, I.-L.\ Lamm, P.\ {M\"{o}ller}, and B.\ Nilsson, Nucl.\ 
  Phys.\ {\bf A131} (1969) 1.                                                   
                                                                                
\item                                                                           
J.\ R.\ Nix, Ann.\ Rev.\ Nucl.\ Sci.\ {\bf 22} (1972) 65.                       
                                                                                
\item                                                                           
M.\ Brack, J.\ Damgaard, A.\ S.\ Jensen, H.\ C.\ Pauli, V.\ M.\ Strutinsky, and 
  C.\ Y.\ Wong, Rev.\ Mod.\ Phys.\ {\bf 44} (1972) 185.                         
                                                                                
\item                                                                           
P.\ M{\"{o}}ller, J.\ R.\ Nix, W.\ D.\ Myers, and W.\ J.\ Swiatecki, {Atomic    
  Data Nucl.\ Data Tables} {\bf 59} (1995) 185.                                 
                                                                                
\item                                                                           
J.\ Duflo, Nucl. Phys. {\bf A576} (1994) 29.                                    
                                                                                
\item                                                                           
X.-L.\ Han, C.-L.\ Wu, D.\ H.\ Feng, and M.\ W.\ Guidry, Phys.\ Rev.\ {\bf C45} 
  (1992) 1127.                                                                  
                                                                                
\item                                                                           
M.\ Bolsterli, E.\ O.\ Fiset, J.\ R.\ Nix, and J.\ L.\ Norton, Phys.\ Rev.\     
  {\bf C5} (1972) 1050.                                                         
                                                                                
\item                                                                           
P.\ {M\"{o}ller} and J.\ R.\ Nix, Nucl.\ Phys.\ {\bf A229} (1974) 269.          
                                                                                
\item                                                                           
P.\ {M\"{o}ller} and J.\ R.\ Nix, Nucl.\ Phys.\ {\bf A361} (1981) 117.          
                                                                                
\item                                                                           
P.\ {M\"{o}ller} and J.\ R.\ Nix, {Atomic Data Nucl.\ Data Tables} {\bf 26}     
  (1981) 165.                                                                   
                                                                                
\item                                                                           
P.\ M{\"{o}}ller and J.\ R.\ Nix, J.\ Phys.\ G: Nucl.\ Part.\ Phys.\ {\bf 20}   
  (1994) 1681.                                                                  
                                                                                
\item                                                                           
W.\ J.\ Swiatecki, Phys.\ Rev.\ {\bf 100} (1955) 937.                           
                                                                                
\item                                                                           
J.\ Randrup, C.\ F.\ Tsang, P.\ {M\"{o}ller}, S.\ G.\ Nilsson, and S.\ E.\      
  Larsson, Nucl.\ Phys.\ {\bf A217} (1973) 221.                                 
                                                                                
\item                                                                           
J.\ Randrup, S.\ E.\ Larsson, P.\ {M\"{o}ller}, S.\ G.\ Nilsson, K.\ Pomorski,  
  and A.\ Sobiczewski, Phys.\ Rev.\ {\bf C13} (1976) 229.                       
                                                                                
\item                                                                           
S.\ Hofmann, V.\ Ninov, F.\ P.\ He{\ss}berger, P.\ Armbruster, H.\ Folger, G.\  
  {M\"{u}nzenberg}, H.\ J.\ Sch{\"{o}}tt, A.\ G.\ Popeko, A.\ V.\ Yeremin, S.\  
  Saro, R.\ Janik, and M.\ Leino, Z.\ Phys.\ {\bf A354} (1996) 229.             
                                                                                
\item                                                                           
P.\ M{\"{o}}ller, J.\ R.\ Nix, and K.-L.\ Kratz, {Atomic Data Nucl.\ Data       
  Tables} {\bf 66} (1997) 131.                                                  
                                                                                
\item                                                                           
Y.\ Aboussir, J.\ M.\ Pearson, A.\ K.\ Dutta, and F.\ Tondeur, Atomic Data      
  Nucl.\ Data Tables {\bf 61} (1995) 127.                                       
                                                                                
\item                                                                           
P.\ M{\"{o}}ller, J.\ R.\ Nix, and W.\ J.\ Swiatecki, Nucl.\ Phys.\ {\bf A469}  
  (1987) 1.                                                                     
                                                                                
\item                                                                           
P.\ M{\"{o}}ller, J.\ R.\ Nix, and W.\ J.\ Swiatecki, Nucl.\ Phys.\ {\bf A492}  
  (1989) 349.                                                                   
                                                                                
\item                                                                           
P.\ M{\"{o}}ller and A.\ Iwamoto, Nucl.\ Phys.\ {\bf A575} (1994) 381.          
                                                                                
\item                                                                           
A.\ Iwamoto, P.\ M{\"{o}}ller, J.\ R.\ Nix, and H.\ Sagawa, Nucl.\ Phys.\ {\bf  
  A596} (1996) 329.                                                             
                                                                                
\end{list}                                                                      
 
\end{document}